
\voffset=-1 cm
\font\twelvebf=cmbx10 scaled \magstep1
\hsize 15.2true cm
\vsize 22.0true cm
\voffset 1.5cm
\nopagenumbers
\headline={\ifnum \pageno=1 \hfil \else\hss\tenrm\folio\hss\fi}
\pageno=1

\hfill IUHET--391

\hfill May, 1998

~
\vskip 4cm

 
\noindent{\twelvebf Are there quasistable strange baryons 
with anticharm or antibeauty?}

\bigskip 
\hskip 2cm  D. B. Lichtenberg$^a$

\hskip 2cm Physics Department, Indiana University,
Bloomington, IN 47405, USA
\bigskip

\hskip 2cm

 \vskip 1 cm

\item{} {\bf Abstract.} In some models, exotic baryons
with strangeness and anticharm or antibeauty should
exist and even be stable against strong decay. We
consider the stability of such possible exotic baryons, 
which in the constituent quark picture are 
called pentaquarks (each is composed of four quarks
and an antiquark). 
Our model is based on diquark clustering and
supersymmetry in hadrons, and assumes that the 
spin-dependent
force between quarks arises from one-gluon exchange.
In the model, a pentaquark with strangeness and anticharm
can decay strongly, but an analogous pentaquark with
an antibeauty quark is stable except for weak decay.

\vskip 1cm

$^a$lichten@indiana.edu

\vskip 2cm

\noindent {\bf 1. Introduction}
\bigskip

Some years ago Lipkin [1] and Gignoux et al.\ [2]
suggested that exotic baryons containing strangeness
and anticharm or antibeauty might be stable against
strong decay. We denote these so-called pentaquarks
by $P(qqqs\bar c)$ and $P(qqqs\bar b)$ 
respectively, where $q$ denotes a $u$ or $d$ quark. 
So far, such exotic baryons have not been observed.

Recently, Roncaglia, Predazzi, and I [3] showed that
a model of hadrons with diquark clustering, supersymmetry,
and chromomagnetic spin-dependent forces as ingredients
could be used to obtain
the masses of some exotic hadrons
in terms of the masses of ordinary
mesons and baryons. The advantage of the model
is that all parameters are obtained
from the meson and baryon sectors---{\it the exotic
hadron masses are obtained without any additional
adjustable parameters.} 
The purpose of this paper
is to apply the model to the lowest-mass 
(ground-state) pentaquarks of the type $P(qqqs\bar c)$
and $P(qqqs\bar b)$. 
(A strange quark has strangeness $S= -1$,
a charmed quark has charm $C = 1$, and a beauty
quark has beauty $B= -1$; antiquarks have the opposite
quantum numbers. By anticharm we mean $C=-1$, but by
antibeauty we mean $B=1$.)


Various ingredients of the model have been developed
in a number of papers. In Section 2 we summarize
the model and give references to papers
where further details may be obtained.
In Section 3 we apply
the model to the pentaquarks $P(qqqs\bar c)$ and
$P(qqqs\bar b)$. We find that the pentaquark
with anticharm
is unstable against strong decay into a nucleon plus
a $D_s$ meson, whereas the  pentaquark
with antibeauty is stable against strong decay. 
It is interesting that
application of heavy quark effective
theory [4] to leading order (neglecting terms that 
go inversely with the heavy quark mass) gives either the 
prediction that these pentaquarks
are stable or that both are unstable. 
We discuss why our model gives a different result, and
we explain why we think that heavy quark effective
theory is not applicable. Finally, we discuss some
of the possible weak decay modes of the pentaquark 
with antibeauty.

 \bigskip
\noindent {\bf 2. Outline of the model}
\bigskip

In this section, we summarize
the basic ingredients of a model incorporating diquark
clustering, hadron supersymmetry, and spin-dependent
forces arising from one-gluon exchange. 
In any hadron containing at least three
quarks (including antiquarks), 
we consider its constituent quarks as far
as possible in pairs, which we call diquarks. 
(A quark and an antiquark is not a diquark.) For example,
a baryon composed of three quarks is considered as
a bound state of a quark and a diquark, an exotic 
four-quark meson (a tetraquark)
is considered as a bound state of a diquark and 
antidiquark,  a pentaquark is a bound state of 
two diquarks and an antiquark, etc. A review of diquark
models has been given by Anselmino et al.\ [5].	

Hadron supersymmetry is an approximate consequence
of QCD. According to QCD, the interaction between two 
coloured particles depends primarily on their individual 
colours and on their colour configuration.  
We now note that a diquark belongs either to a {\bf 6}
or to a ${\bf \bar 3}$ representation of colour-SU(3).
The ${\bf \bar 3}$ is the same representation as
that of an antiquark.  Therefore, the transformation of
replacing an antiquark (a fermion) by a  ${\bf \bar 3}$
diquark (a boson) in a hadron should to first approximation
leave the interaction energy unchanged. This is what
we mean by hadron supersymmetry. The mathematical
underpinnings of the idea can be found in papers
of Miyazawa [6] and Catto and G\"ursey [7]. 

A limitation of hadron supersymmetry 
is that although the transformation is an approximate
invariant for ${\bf \bar 3}$ diquarks, the interactions of
sextet diquarks are not 
related to those of antiquarks in the model.
However, in lowest-order perturbation theory, the 
interaction of two quarks in a colour-${\bf \bar 3}$
state is attractive, while in a colour-{\bf 6} state 
the interaction
is repulsive. This fact leads us to conjecture that
sextet diquarks have higher masses than antitriplet
diquarks with the same quark content, so that if we are 
considering low-mass states
we can safely neglect configurations containing sextet
diquarks.

The supersymmetry is broken by mass, spin, and size
differences between an antiquark and a diquark. 
Noting that neither a constituent quark nor a diquark
is a point particle, we neglect the size effect. 
However, we are able to make approximate corrections for 
differences in mass and spin. Let us first consider spin.

We approximately remove the effect of spin-dependent forces
by averaging over spins of
mesons and baryons. We confine ourselves to ground
states, and and confine our analysis to states
with zero orbital angular momentum in the nonrelativistic
approximation.  The 
spin-averaged mass ${\cal M}$ of a vector meson $M_1$ and 
pseudoscalar meson $M_0$ is given by [8]
$${\cal M}=(3M_1 + M_0)/4. \eqno(1)$$
We can use this simple formula to obtain the value of
${\cal M}$ when $M_1$ and $M_0$ are known from 
experiment [9]
In cases in which data are lacking, we can obtain 
estimates by interpolation or extrapolation,
making use of empirical regularities in hadron masses
[10, 11].  The meson 
interaction energy $E_{\cal M}$ is defined in terms of the
mass ${\cal M}$ and  the constituent quark masses $m_1$
and $m_2$:
$$E_{\cal M} ={\cal M}- m_1 -m_2. \eqno(2)$$
We denote the spin-dependent part of the interaction energy 
between quark and antiquark by  $\epsilon$, given by
$$ \epsilon = (M_1 - M_0)/4. \eqno(3)$$
Then 
$$M_1 = {\cal M} + \epsilon, \quad M_0 =
{\cal M} - 3\epsilon. \eqno(4)$$

For mesons, this procedure is straightforward, but for
baryons we need a model to extract the spin-dependent
interaction energy between two quarks from baryon
mass differences, given that a baryon contains three
quarks. We assume that the spin-spin interaction is the 
chromomagnetic interaction arising from one-gluon exchange. 
A different assumption will in general give different
answers. In order to write down results, we have to
order the three quarks in a baryon. We adopt the
following ordering: if all three quarks 
have different flavours, we order them from lightest
to heaviest; if any two have the same flavour, they are
the first two [8].
In the case in which all three quarks 
have different flavours, there are three ground-state
baryons, which all have zero orbital angular momentum
but differ in their spin configurations.
Spin-dependent forces break the degeneracy.
We denote by $A$ and $A'$ the baryons of spin 1/2, 
which differ according to whether the first two quarks
have spin zero or one. We denote by $A^*$ the baryon
of spin 3/2. (In the case of baryons containing
two $q$ quarks and an $s$ quark, 
$A$ is  $\Lambda$, $A'$ is $\Sigma$, and $A^*$ is 
$\Sigma^*$.) Then the spin-averaged mass ${\cal A}$
and interaction energy $E_{\cal A}$ are
given by [8]
$${\cal A}= (2 A^* + A + A')/4, \eqno(5)$$
$$E_{\cal A} = {\cal A} - m_1 -m_2 -m_3. \eqno(6)$$
We let $\epsilon_{12}$, $\epsilon_{13}$, and 
$\epsilon_{23}$ be given by
$$ \epsilon_{12}= (2A^* + A' - 3A)/12, \quad 
\epsilon_{13}+ \epsilon_{23}= (A^* - A')/3. \eqno(7)$$
Then
$$A^* = {\cal A} + \epsilon_{12}+
\epsilon_{13} + \epsilon_{23}, \quad
A' = {\cal A}+ \epsilon_{12}-
2\epsilon_{13} - 2\epsilon_{23}, \quad
A = {\cal A}- 3\epsilon_{12}. \eqno(8)$$
The values of $\epsilon_{ij}$ turn out to depend weakly 
on the mass of the ``spectator'' quark $k$.

If any two quarks have the same flavour, the baryon $A$
is absent because of the Pauli principle, while if all 
three quarks have the same 
flavour, both $A$ and $A'$ are absent.
We see from Eq. (8)  that in these cases
we cannot in general determine ${\cal A}$ from the
baryon masses by spin averaging. However, 
we can estimate the values of $\epsilon_{ij}$
from empirical regularities [10, 11], and then
obtain ${\cal A}$ from Eq. (8).

We can obtain some of the values of 
${\cal M}$ and ${\cal A}$
in terms of the masses of observed mesons and baryons [9],
and we can determine the others with the help of
observed regularities in hadron masses [10, 11]. 

The constituent quark masses in Eqs. (2) and (6)
are not completely 
arbitrary, but satisfy certain constraints [11, 12].
For definiteness, we use the values obtained in Ref.\ [12].
They are (in MeV)
$$m_u=m_d= 300, \quad m_s = 475, \quad m_c =1640,
\quad m_b= 4985. \eqno(9)$$
It is important to note that these quark masses were
obtained using as input only normal meson and baryon 
masses. No input from exotic hadrons was used.

Using these quark masses, we can determine $E_{\cal M}$
and $E_{\cal A}$ from the values of 
${\cal M}$ and ${\cal A}$.
If we now plot $E_{\cal M}$ as a function of the reduced
mass $\mu$ of its constituent quarks, we find the function
is rather smooth and suitable for interpolation. 
Because of supersymmetry,
if we can estimate the mass of a diquark, we can regard
it as a fictitious antiquark of the same mass, and
find the interaction energy from the meson curve of
$E_{\cal M}$ vs the reduced mass $\mu$. 

We estimate the diquark masses as follows: We choose
the two heaviest quarks in a given baryon to be the 
diquark. We guess that its mass is the sum of its
constituent masses, calculate the reduced mass of
diquark and the third quark, and read off the 
interaction energy from the curve for mesons. Adding
this interaction energy to the mass of the quark and
diquark will give a mass which will not in general have
the value ${\cal A}$. However, we then vary the diquark
mass until the baryon mass obtained as the sum of quark
and diquark masses plus the interaction energy from the
meson curve gives the value ${\cal A}$. This diquark
mass ${\cal D}$ is the spin-averaged diquark mass.
Then from the formula
$$E_{\cal D} = {\cal D} - m_i -m_j, \eqno(10)$$
we obtain the interaction energy of of two colour-triplet
particles to form a colour antitriplet. Again it turns
out that the values of $E_{\cal D}$ are rather smooth
functions of $\mu$.

We next assume that the spin-dependent force between
two quarks in a diquark is the same as the value
we found for the diquark in a baryon. (We must neglect
the weak dependence on the spectator quark.) Then 
we obtain for the masses of spin-one and spin-zero
diquarks containing quarks $i$ and $j$
$$D_1 = {\cal D} + \epsilon_{ij}, \quad
D_0 = {\cal D} - 3\epsilon_{ij}. \eqno(11)$$
The values of $D_0$ and $D_1$ are given in 
Table 1 of Ref.\ [3].

\bigskip
\noindent {\bf 3. Results and discussion}
\bigskip

Let us first consider the pentaquark $P(qqqs\bar c)$, which 
in our model contains diquarks made of $qq$ and $qs$. 
Because we are interested in the pentaquark of lowest
mass, we choose both these diquarks 
to have spin zero, as the chromomagnetic interaction
causes spin-zero diquarks to
have lower mass than those of spin one. Note that
the $qq$ diquark must be $ud$ to satisfy the Pauli
principle. 
We neglect the Pauli principle
for identical quarks in different diquarks, just as in
nuclear physics  the Pauli principle is usually
neglected for quarks
in different nucleons. This approximation is probably
better for nucleons than for diquarks. A brief discussion of
this approximation is given in Ref.\ [3]. 

   From Ref.\ [3] the $qq$ and $qs$ diquarks have 
masses given by (in MeV): 
$$D(qq) = 595, \quad D(qs) = 835. \eqno(12)$$
Because there are no spin-dependent forces between
spin-0 particles, we can directly read off from a
curve giving the interaction energy of two bound triplets
as a function of $\mu$
that this interaction energy $E_Q$ ($Q$ stands for
quadriquark) is
$$E_Q = 170 \ {\rm MeV}. \eqno(13)$$
We then obtain that the spin-zero quadriquark mass $Q$ is 
$$ Q= 1600 \ {\rm MeV}. \eqno(14)$$
We calculate the reduced mass of the colour triplet
(the quadriquark) and the colour antitriplet
(the $\bar c$ quark), and then read off the interaction
energy from the meson curve giving the energy vs 
reduced mass. We obtain
$$\mu = 810\ {\rm MeV}, \quad E_P =-210\ {\rm MeV}.
\eqno (15)$$
Adding the masses of the quadriquark and charmed quark
to the value of $E_P$, we obtain that the mass of the 
penatquark is
$$P(qqqs\bar c)= 3030 \ {\rm MeV}. \eqno(16)$$
We see that with this mass the anticharmed pentaquark
can decay strongly as follows:
$$P(qqqs \bar c) \rightarrow N(938) + \bar D_s(1969)
+ 123 \ {\rm MeV}. \eqno(17)$$

If we go through the same steps for the pentaquark
$P(qqqs\bar b)$, we find that the reduced mass 
and interaction energy are
$$\mu = 1210\ {\rm MeV}, \quad E_P = -320\ {\rm MeV}. 
\eqno(18)$$
This leads to a mass value
$$P(qqqs\bar b)= 6265 \ {\rm MeV}. \eqno(19)$$
But the lowest mass state of a normal meson and baryon
with the same quantum numbers as the $P(qqqs\bar b)$
is $N(938) + B_s(5369)$, which has a total mass
of 6307 MeV, a value {\it greater} than the mass of 
the $P(qqqs\bar b)$. Consequently, in our model
the pentaquark with an antibeauty quark 
cannot decay strongly. 

By construction, these ground-state pentaquarks have
no orbital angular momenta.  It follows that each
has spin-parity $J^P = {1\over 2}^-$ (the negative
parity arises from the fact that an antiquark has
negative intrinsic parity).

Let us apply heavy quark effective theory [4] 
in leading order
to Eq.\ (17) by replacing the $\bar c$ quark with
a $\bar b$ quark. Then the $P(qqqs\bar c)$ becomes
$P(qqqs\bar b)$ and $\bar D_s$ ($\bar c s$) becomes $B_s$
($\bar b s$). But according to 
heavy quark effective theory
(neglecting terms in $1/m_c$ and $1/m_b$), the hadron 
masses satisfy 
$$P(qqqs\bar b) - P(qqqs\bar c) \simeq
B_s - D_s \simeq m_b - m_c, \eqno(20)$$
so that Eq.\ (17) becomes
$P(qqqs \bar b) \rightarrow N(938) + \bar B_s(5369)
+ 123 \ {\rm MeV}. $ From these numbers, 
the mass of $P(qqqs\bar b)$
is 6430 MeV, rather than our result of 6265 MeV. 
The difference between the two mass predictions is only 
165 MeV, or about 2.6\% of the $P(qqqs \bar b)$ mass. 
Of course, this prediction of the mass of the
antibeauty pentaquark from heavy quark
effective theory depends on our result
for the mass of $P(qqqs \bar c)$. Without that input, 
we obtain only Eq.\ (20) from heavy quark effective theory.

Wherein lies the difference between the predictions
of our model and those of 
heavy quark effective theory in leading order? 
According to the latter
theory, both the $c$ and $b$ quarks are considered
infinitely heavy compared to the masses of the light
quarks. This might be a good approximation for the
$D_s$ meson, since the 1640 MeV mass of the $c$
quark is considerably greater than the 475 MeV mass of
the $s$ quark. However, the approximation breaks down
for the $P(qqqs\bar c)$ because even though the mass
of the $\bar c$ quark is much greater than the mass of
any of the other quarks in this hadron, the $\bar c$-quark
mass is comparable to the {\it sum} of the masses
of the light quarks, i.e., to the mass of the quadriquark
$Q=qqqs$.  As a result,
the reduced mass of the quadriquark and $\bar c$ quark
is considerably less than the mass of the quadriquark.
On the other hand, the $\bar b$ quark is much heavier than
the quadriquark so that the reduced mass of those
two particles is not much smaller than the quadriquark
mass. Because  the interaction energy 
decreases as the value of $\mu$ increases [11], the
$\bar b$ pentaquark is more tightly bound than the 
$\bar c$ pentaquark, resulting in  the former being stable
against strong decay.

If the antibeauty pentaquark decays weakly, its lifetime
should be similar to the lifetimes of observed
$b$-quark hadrons, or 1 to 1.5 $\times 10^{-12}$ s. 
It comes in two charge states,
$$P^+ = P(uuds\bar b), \quad P^0 = P(udds\bar b) \eqno(21)$$
Decay modes should include one hadron containing a 
$\bar c$ quark. Prominent decay modes should include
$$P\rightarrow N + \bar D_s\ ( + \ {\rm pions}),$$
$$P \rightarrow \Lambda \ ({\rm or}\ \Sigma) + \bar D
\ (+ \ {\rm pions}), \eqno(22)$$
$$P \rightarrow N + \bar K + \bar D \ (+ \ {\rm pions}). $$
There should also be decays with both hadrons and leptons
in the final state.

It is interesting that in the model the $P^0$ has a 
possible two-body decay
mode into a nucleon and a meson:
$$P^0 \rightarrow p + \bar D_s, \eqno(23)$$
whereas the simplest two-body decay mode of the $P^+$
leads to a $\Sigma^+$ baryon:
$$P^+ \rightarrow \Sigma^+ + \bar D^0. \eqno(24)$$
However, we expect that most of the decays will 
be into multibody final states.

In our model, the pentaquark $P(qqqc\bar s)$ 
has a higher mass
than $P(qqqs\bar c)$. Furthermore, the lowest-mass state
with the same quantum numbers as
$P(qqqc\bar s)$ is $\Lambda_c(2285) +
K(496)$, considerably less massive than the state
$N(938) + \bar D_s(1969)$ into which the $P(qqqs\bar c)$
can decay. Consequently, we predict that the $P(qqqc\bar s)$
will have a large decay width, which, together with a small
production cross section, will 
make it unlikely to be observed.
For similar reasons we predict that $P(qqqb\bar s)$ will
not be observed.

We emphasize that our model depends on three 
ingredients: diquark clustering, hadron supersymmetry,
and spin-dependent forces
arising from the chromomagnetic interaction of one-gluon
exchange. With these simple inputs, we have been able
to calculate the masses of exotic baryons with no
adjustable parameters---all parameters having been fixed
from the observed properties of ordinary mesons and baryons. 

But of course, there are limitations to the model.
It is possible that the actual structure
of exotic hadrons is different from the one we have
assumed. Furthermore, if diquark clustering is important,
a small admixture of colour-sextet diquarks might lower the
pentaquark masses. Also, the spin-dependent forces
might arise from a mechanism other than from 
chromomagnetism. For example, 
it has been proposed [13] that the spin-dependent
interaction among light quarks arises from the
breaking of chiral symmetry by one-boson exchange. 
Such a mechanism would lead to a quantitative change
in our results
and might also alter our qualitative conclusions.

In our approach we do not use any explicit Hamiltonian.
Recently, pentaquarks have been considered in a
conventional Hamiltonian model [14, 15], with 
spin-dependent forces arising from chiral symmetry breaking.
The Hamiltonian model is quite different from our
picture, and leads to different results.

Of course, it is only experiment that
can distinguish between models.
Therefore, experimentalists should search for both
the anticharmed pentaquark and for the pentaquark
with antibeauty. Moinester et al.\ [16] have discussed
ways to look for the former. A search for the latter
will be much more difficult, but in view of our
prediction that it is quasistable, a serious effort
to find it should be undertaken.

\bigskip
\noindent {\bf Acknowledgments}
\bigskip

I should like to thank Steven Gottlieb, Renato
Roncaglia, and Floarea Stancu for  helpful
discussions.  This work was supported in
part by the U.S. Department of Energy. 

\vfill\eject

\vskip 1cm
\noindent {\bf References}
\medskip
\font \eightrm=cmr9
\par
\everypar{\parindent=0pt \hangafter=1\hangindent=2 pc}
 
\eightrm
\par

\noindent [1] Lipkin H J 1987 {\it Phys.\ Lett.}\ {\bf B195}
484 

\quad \  Lipkin H J 1988 {\it Nucl.\ Phys.}\ {\bf A475} 307c

[2] Gignoux C, Silvestre-Brac B and Richard J M
1987 {\it Phys.\ Lett.}\ {\bf B193}

[3] Lichtenberg D B, Roncaglia R and Predazzi E 1997 
{\it J.\ Phys.\ G: Nucl.\ Part.\ Phys.}\ {\bf 23} 865

[4] Isgur N and Wise M B 1989 {\it Phys.\ Lett.}\ {\bf
B232} 113

[5] M Anselmino M,  Predazzi E,  Ekelin S,
Fredriksson S and  Lichtenberg D B 1993
{\it Rev.\ Mod.\ Phys.}\ {\bf 65} 1199 

[6] Miyazawa H 1966  {\it Prog.\ Theor.\ Phys.}\
{\bf 36} 1266

\quad \  Miyazawa H 1968 {\it Phys.\ Rev.\ }
{\bf 170} 1586

[7] Catto  S and  G\"ursey F 1985
{\it Nuovo Cimento} {\bf 86} 201

\quad \ Catto  S and  G\"ursey F 1988
{\it Nuovo Cimento} {\bf 99} 685

[8]  Anselmino M,  Lichtenberg D B and
Predazzi  E 1990 {\it Z.\ Phys.\ C}
{\bf 48} 605

[9] Particle Data Group: Barnett R M et al. 1996,
{\it Phys.\ Rev.\ D} {\bf 54} 1

[10] Song X 1989 {\it Phys.\ Rev.\ D} {\bf 40} 3655 

\quad \ Wang Y and  Lichtenberg D B 1990
{\it Phys.\ Rev.\ D} {\bf 42}  2404

[11]  Roncaglia R,  Dzierba A R,  Lichtenberg D B,
and Predazzi E 1995, {\it Phys.\ Rev.\ D} {\bf 51}  1248

[12] Roncaglia R,   Lichtenberg D B,
and  Predazzi E 1995 {\it Phys.\ Rev.\ D} {\bf 52}  1722

[13] Glozman L Ya and Riska D O 1996 {\it Phys.\ Rep.}\
{\bf 268} 264

[14] Genovese M, Richard J-M, Stancu
Fl and Pepin S, e-print hep-ph/9712452,
{\it Phys.\ Lett.\ B}, to be published

[15] Stancu Fl, unpublished e-print

[16] Moinester M A, Ashery D, Landsberg L G and Lipkin
H J 1996 {\it Zeit.\ Phys.\ A} {\bf 356} 207

\bye